\documentclass[pre, aps, amssymb,amsmath,showpacs,nofootinbib,floatfix,twocolumn]{revtex4}

\usepackage{graphicx}
\usepackage{epsfig}

\newcommand{\CM}{{\mathbb C}}
\newcommand{\ZM}{{\mathbb Z}}
\newcommand{\RM}{{\mathbb R}}

\newcommand{\TM}{{\mathbb T}}

\newcommand{\rot}{\hat{\cal R}_{\tau}}

\newcommand{\map}{{\cal U}_{\alpha,\tau}}

\newcommand{\mapth}{{U}_{\alpha}}

\newcommand{\IPR}{\small\texttt{I}}

\begin{document}
\title{The Gross-Pitaevski map as a chaotic  dynamical system}

\author{Italo Guarneri$^{1,2}$}
\affiliation {{\small
$^1$ Center for Nonlinear and Complex Systems}\\
 { Dipartimento di Scienza ed Alta Tecnologia - Universit\'a dell'Insubria, via Valleggio 11, I-22100 Como, Italy.}\\
 {\small $^2$ Istituto Nazionale di Fisica Nucleare, Sezione di Pavia,
via Bassi 6, I-27100 Pavia, Italy.}}

\begin{abstract}
{The Gross-Pitaevski map is a discrete time, split-operator version of the Gross-Pitaevski dynamics in the circle,
for which exponential instability has been recently reported. Here it is studied as a classical dynamical system in its own right. A systematic analysis of Lyapunov exponents exposes strongly chaotic behavior. Exponential growth of energy is then shown to be a direct  consequence of rotational invariance and for stationary solutions the full spectrum of Lyapunov exponents is analytically computed. The present analysis includes the "resonant" case, when the free rotation  period is commensurate to $2\pi$, and the map has countably many constants of the motion. Except for lowest order resonances, this case exhibits an integrable-chaotic transition. }
\end{abstract}
\pacs{05.45.-a,03.75.-b}
\maketitle

\section{Introduction}

 A few nonlinear variants  of the quantum kicked rotor  have been devised \cite{dim,gp,raiz, gc} in order to investigate the impact of nonlinearity on the dynamical localization, which is the prototypical feature of that model \cite{kr}.  For some of them  possible experimental realizations with Bose-Einstein condensates have been surmised.  The most recent \cite{gp,gc} is described  by a family of nonlinear maps in the $\infty$-dimensional space of functions $\psi(\theta)$ on the $1-$ torus $\TM$:
\begin{equation}
\label{map0}
{\cal U}_{\alpha,\tau,k}\;=\: {\cal V}_{\alpha,k}\circ{\rot}\;,
\end{equation}
where $\rot$ is the linear operator $\exp(i\tfrac{\tau}{2}d^2/d\theta^2)$ , and ${\cal V}_{\alpha,k}$ is the  nonlinear operator:
\begin{equation}
\label{map}
{\cal V}_{\alpha,\tau}(\psi)(\theta)\;=\;e^{i\alpha\psi(\theta)^*\psi(\theta)\;+\;ik\cos(\theta)}\;\psi(\theta)\;.
\end{equation}
$k,\tau,\alpha $, are real parameters, with $k\geq 0$, $\tau\geq 0$.  The linear quantum kicked rotor is described by ${\cal U}_{0,k,\tau}$.
Map (\ref{map0}) is also obtained when a split-operator method \cite{DT} is used in its simplest version to approximate the continuous-time Gross-Pitaevski (GP) (or "cubic" nonlinear  Schr\"odinger) dynamics on the torus. For this reason it will be dubbed GP map in the following. However here it is studied  as a dynamical system in its own right,  and not as an accessory of the standard GP equation.
In particular,  chaotic motion in the GP map is studied. Motivation is provided by reports \cite{gp,gc} of   exponential growth of energy (energy in state $\psi$ is defined as $E(\psi)=\tfrac12\|\psi'\|^2$, where $'$ denotes $\theta$-derivative), and exponentially fast separation of wave packets \cite{gc} when the period $\tau$ is incommensurate to $2\pi$. In the present paper such issues are investigated in detail, and the above results are completed and significantly extended. The onset of chaos is studied by a systematic analysis of Lyapunov exponents (LE), computed by numerically iterating the tangent map to (\ref{map0}).   As exponential instability is not crucially related to the presence of a cosine potential, the suffix $k$ is henceforth removed and $k=0$ is understood. \\ Exponential growth of energy is just a special case of this Lyapunov analysis, because it is equivalent to exponentially fast divergence of trajectories which initially differ by an infinitesimal rotation. It is worth noting that  early studies on the "quantum suppression of classical chaos"  have long ago pointed out that quantum chaotic behaviour - if any! - should  display exponential energy growth \cite{shp}. Anyway in the present paper the GP map is a classical dynamical system; a comment about quantum chaos is deferred to the concluding remarks in sec.\ref{concr}.\\
The alleged exponential instability of $\infty$-dimensional dynamics poses a delicate task to numerical investigation, which of necessity uses  finite dimensional approximations. Such approximations  define  measure preserving  dynamical systems on finite dimensional phase spaces, for which - unlike the $\infty$-dimensional case - existence of Lyapunov exponents is an exact result; it is these very  LEs that are numerically computed, finding evidence of chaotic transitions in the finite dimensional dynamics.
Remarkably, they are observed to uniformly stabilize  when the basis size is large enough, supporting the conjecture, that they reflect properties of the $\infty$-dimensional dynamics.
A theoretical explanation is obtained, noting that, for large dimension,  ergodicity and Levy's lemma about  the concentration of measure \cite{conc} justify a mean-field approach to the tangent dynamics.  Doing so allows a transparent derivation of a simple formula, that describes the  dependence of the maximal Lyapunov exponent on the nonlinearity parameter $\alpha$. The finite dimension does not appear any more in this formula, which coincides with one  that  was obtained in \cite{gc} by means of a different argument and for  the special case of energy growth. For "stationary" orbits of the GP map the whole Lyapunov spectrum is analytically computed in sect.\ref{sstat}; it turns out that, whenever the period $\tau$ is incommensurate to $2\pi$, and  for any nonvanishing nonlinearity, such orbits are linearly unstable in infinitely many directions, and linearly stable in infinitely many ones. While individual LEs depend on the specific value of $\tau$, their distribution does not.

The sign of $\alpha$, which  plays a crucial role in the continuous-time GP dynamics as it discriminates between attractive ($\alpha>0$ ) and
repulsive ($\alpha<0$) interaction, is substantially irrelevant for the aspects of the GP map which are discussed in the present paper. One reason for this difference is that the map does not conserve the total GP energy $E(\psi)-\tfrac12 \alpha \|\psi^2\|^2$.
The main general features of the observed exponential instability depend on $\tau$ only through its commensurate/incommensurate (to $2\pi$) character. The hitherto unexplored commensurate case, when $\tau=4\pi P/Q$, with $P,Q$ mutually prime integers, is studied in Sect.\ref{reso}. In that case the GP map has infinitely many independent constants of the motion and the $\infty$-dimensional phase space is fibered in $Q$-dimensional fibers. Over each such fiber the GP map defines  a (Lebesgue) measure-preserving dynamical system, with well-defined  Lyapunov exponents. This system  appears to undergo an integrable-chaotic transition as $|\alpha|$ is increased, that is also  mirrored in the global dynamics.

\section{Basic properties}
\label{basic}
  The  following properties of the GP map are straightforward:\\
  1 - $\map$ is a continuous, invertible operator in  the Hilbert space $L^2({\TM})$. It also preserves the Hilbert  norm : $\int_0^{2\pi}|\map(\psi)(\theta)|^2d\theta=\int_0^{2\pi}|\psi(\theta)|^2d\theta$. Thanks to this property, one may always resort to states normalized to unity by rescaling $\alpha$. Such normalization is assumed throughout the following. $\map$ is continuous, though not  norm-preserving, also in the Hilbert space $H_1(\TM)$ of absolutely continuous functions with a square-integrable derivative, with the "energy norm":
  $$
    \|\psi\|_1^2\;=\;\int_0^{2\pi}d\theta\;\left(|\psi(\theta)|^2\;+\;|\psi'(\theta)|^2\right)\;
  $$
  \noindent
  2 - $\map$  preserves a symplectic form, see eqn.(\ref{symp}) below; \\
  3 - $\map$  is rotation invariant: $\forall\eta\in\RM$, $\map({\hat T}_{\eta}\psi)={\hat T}_{\eta}\;\map(\psi)$, where ${\hat T}_{\eta}\psi(\theta)=\psi(\theta-\eta)$. \\
  4 - scaling symmetry : $\forall\gamma\in\CM$, $\map(\gamma\psi)=\gamma\;{\cal U}_{|\gamma|^2\alpha, \tau}(\psi)$ \\
  5 - whenever $\tau$ is commensurate to $2\pi$: $\tau=4\pi P/Q$, $P,Q$ integers, $\map(e^{iQ\theta}\psi)=e^{iQ\theta}\map(\psi)$.\\
  6 - analyticity is preserved: if $f(z)$ is an analytic function in $\Omega_R=\{R^{-1}<|z|<R\}$ and $\psi(\theta)=f(e^{i\theta})$, then $\map(\psi)$ is   analytically continued in $\Omega_R$ to :
  $$
  \map(f)(z):=e^{i\alpha {\rot f}(z)({\rot f})^*(1/z^*)}{\rot}f(z).
  $$
  This in particular implies that the asymptotic rate of exponential decay of $\hat{\psi}$ over the Fourier basis  is unchanged under the GP dynamics.

\section{Lyapunov exponents.}
\label{lexp}

The tangent dynamics to the map (\ref{map}) will be studied in the {\it real} Hilbert  spaces $\cal H$ , ${\cal H}_1$ of couples $(\psi, \psi^*)$, with $\psi\in L^2({\TM})$ or $\psi
\in H_1$ respectively .    Formal differentiation
with respect to $\psi$, $\psi^*$ along  a trajectory $\{\psi_t\}_{t\in\ZM}$ , $\psi_t=\map^t(\psi_0)$, yields the following "variation equation":
\begin{equation}
\label{var}
\begin{split}
\delta\psi_{t+1}(\theta)\;&=\;e^{i\alpha{\tilde \psi}_t(\theta)^*{\tilde\psi}_t(\theta)}\biggl(
\delta{\tilde\psi}_t(\theta)\;
+\;\\
&2i\alpha{\tilde\psi}_t(\theta)\text{\rm Re}\left({\tilde\psi}_t(\theta)^*\delta{\tilde\psi}_t(\theta)\right)\biggr)\;,
\end{split}
\end{equation}
where:
\begin{equation}
{\tilde\psi}_t(\theta)\;=\;{\rot}{\psi}_t(\theta)\;,\;\;\;\;\;\delta{\tilde\psi}_t(\theta)\;=\;{\rot}\delta{\psi}_t(\theta)\;.
\end{equation}
Eqn.(\ref{var}) defines  linear operators ${\cal T}_{{\psi}}$ in $\cal H$, ${\cal H}_1$ such that $\delta\psi_{t+1}={\cal T}_{\psi_t}(\delta\psi_t)$ (where $\delta\psi_t$ now stands for $(\delta\psi_t,\delta\psi_t^*)$, and similarly for $\psi_t$).  They are not bounded in
$\cal H$ except for special choices of $\psi$. In ${\cal H}_1$
the operator ${\cal T}_{\psi}$ is the Fr\'echet differential of the map that is defined in ${\cal H}_1$ by (\ref{map0}). It preserves the symplectic form:
 \begin{equation}
 \label{symp}
\sigma(\delta\psi,\delta\phi)\;=\;\text{\rm Im}\langle \delta\psi|\;\delta\phi\rangle
\;=\;\text{\rm Im}\int_0^{2\pi}d\theta\;\delta\psi(\theta)^*\delta\phi(\theta)\;,
\end{equation}
or, in the Fourier basis :
\begin{equation}
\label{symp0}
\sigma(\delta\psi,\delta\phi)\;=\;\sum\limits_{n\in\ZM} \text{\rm Im}\left(\widehat{\delta\psi}(n)^*\widehat{\delta\phi}(n)\right)\;.
\end{equation}
Exponential instability of a trajectory $\{\psi_t\}_{t\in\ZM}$ is  related to positivity of some Lyapunov exponent (LE)  of the trajectory. Here LEs are defined by:
\begin{equation}
\label{le1}
\Lambda\;=\;\limsup\limits_{t\to\infty}\tfrac 1t\lambda_t\;,\;\;\;\;\;\
\lambda_t\;=\;\log\left(\|\delta\psi_t\|/\|\delta\psi_0\|\right)\;.
\end{equation}
with $\delta\psi_t={\cal T}_{\psi_{t-1}}\circ{\cal T}_{\psi_{t-2}}\circ\ldots\circ{\cal T}_{\psi_0}(\delta\psi_0)$.
\begin{figure}
\begin{center}
\epsfxsize8cm\epsffile{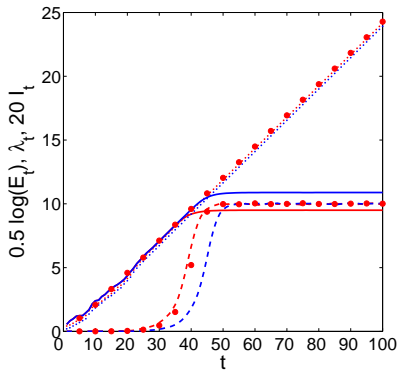}
\end{center}
\caption{\scriptsize Growth with time $t$ of the logarithm of energy (full lines), of $\lambda_t$ (eqn.(\ref{le1})) (dotted lines, pluses) , and of the normalized inverse participation ratio (eqn.(\ref{ipr}))(dashed lines), for $\tau=2\pi(\sqrt{5}-1)$, and with $\psi_0$ and $\delta\psi_0$ as specified in the text; $\alpha=2.5$ (lines), $\alpha=-2.5$ (circles). Fourier basis size: $2^{17}+1$ (red), $2^{18}+1$ (blue).}
\label{vstime}
\end{figure}
\begin{figure}
\begin{center}
%\vskip 0.2cm
\epsfxsize5cm\epsffile{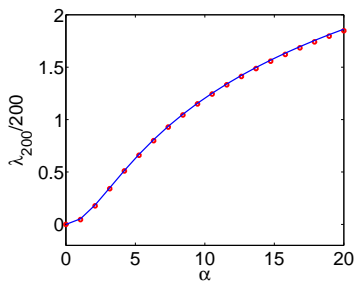}
\end{center}
\caption{\scriptsize Circles: numerically computed finite-time Lyapunov exponent $\lambda_t/t$ for $t=200$ vs $\alpha$, basis size $2^{17}+1$, $\tau=2\pi(\sqrt 5-1)$.  Full line: eqn.(\ref{lyap}).}
\label{vsalpha}
\end{figure}
No argument is given here why such $\Lambda$ should be finite. In principle, they  depend on the chosen trajectory, and on the choice of an initial $\delta\psi_0$. All trajectories exhibit at least one  zero LE. For instance, if $\delta\psi_0=ic\psi_0$ with $c$ real arbitrary, then eqn.(\ref{var}) entails $\delta\psi_t= ic\psi_t$ at all times.

The rate of exponential growth of energy along a trajectory, divided by $2$, is itself a LE.  This follows from:
\begin{gather}
\psi'_{t+1}\;=\;\lim\limits_{\eta\to 0}\frac{\hat T_{\eta}-1}{\eta}\map(\psi_t)\nonumber\\
=\;\lim\limits_{\eta\to 0}\frac{\map(\hat T_{
\eta}\psi_t)-\map(\psi_t)}{\eta}\nonumber\\
=\;{\cal T}_{\psi_t}(\psi'_t)\;,
\label{rotinv1}
\end{gather}
where $'$ denotes $\theta-$derivative. Hence:
\begin{equation}
\label{rotinv2}
\tfrac1t\log(E(\psi_t))\;=\;\tfrac2t\log(\|\delta\psi_t\|)\;+\;O(1/t)\;,\;\;\;\;\;\;\delta\psi_0=\psi'_0\;.
\end{equation}
The present study of LEs is based on numerical solution of the variation equation (\ref{var}). Fig.\ref{vstime} shows the behavior in time  of the logarithm of energy, and  of $\lambda_t$ as defined in (\ref{le1}). The initial $\psi_0$  is a coherent state centered at momentum $0$ and $\theta=0$, with $h=0.1$; $\delta\psi_0$ is a randomly generated vector:
$\hat{\delta\psi_0}(n)=
$rand$(n)\hat{\psi_0}(n)$, with rand$(n)$ independent, normally distributed random variables. Though most numerical results shown  in this paper  were obtained with $\alpha>0$, no substantial difference is observed with $\alpha<0$; see  the data for $\alpha=-2.5$ in fig.\ref{vstime}.
Of course, numerical simulations do not address the true $\infty$-dimensional dynamics, but the finite-dimensional dynamics  that are obtained on restricting $\map$ to a finite-N- dimensional Fourier subspace, and on discretizing the circle in a grid of $N$ points. Such finite-dimensional  dynamics implement finite Fourier transforms, and are defined by  norm-preserving maps in a hypersphere in $\CM^N$, where  they
 preserve the symplectic form
(\ref{symp}). They are therefore (Lebesgue) measure-preserving, and define classical dynamical systems, for which existence (almost everywhere) of LEs is an exact result.  Energy in state $\psi$ is computed as $1/2\sum_{-M}^{M}n^2|{\hat\psi}(n)|^2$, where $N=2M+1$.
It is initially almost constant (fluctuations are suppressed in the  logarithmic scale), then fast exponential growth $\propto\exp(2\Gamma t)$ is observed, quickly leading to saturation of the basis; energy remains thereafter stationary. Instead the "local divergence of trajectories", as  measured by $\lambda_t$, after a short logarithmic increase (the duration of which depends on $\alpha$) enters a steady linear growth, yielding a positive LE $\Lambda$. Sampling different choices of $\psi_0$ suggests that $\Lambda$ should be independent of $\psi_0$ (with at least one notable exception: the Fourier
basis functions studied in Sec.\ref{sstat}, see below).
The finite-N dynamics only enjoys a discrete rotational invariance, so the argument in eqs.(\ref{rotinv1}), (\ref{rotinv2}) breaks down, and energy does not any more define a LE proper. When $\alpha>\sim 1.5$  it was observed that $\Lambda\approx\Gamma$, (half the transient rate of exponential growth of energy), as in eq.(\ref{rotinv2}); not so at smaller $\alpha$, however: {\it e.g.}, for $\alpha=0.4$, $\Gamma\approx 0.1$ and $\Lambda\approx 0.007$ are read in the exponential range. It must be noted, however, that exponential instability is observed the later, the smaller $\alpha$ is, and
 its reliable detection eventually falls beyond the computational capabilities of the present work. At smaller times, $\lambda_t$ increases logarithmically, as  trajectories separate linearly in time \footnote{The small-$\alpha$ region is the relevant region for the split-operator approximation of the continuous time GP dynamics. This suggests that the  validity of that approximation may not be crucially affected by the chaotic behavior of the GP map.
 }

For large basis size $N$ the numerically computed  $\Lambda$ appear to stabilize, independently of
which increasing sequence of $N-$ values is used; {\it e.g.} using the denominators of the principal convergents to $\tau/(2\pi)$ no significant difference is observed. That notwithstanding, identifying them  with LE exponents of the $\infty$-dimensional dynamics implies interchanging the $N\to\infty$ and the $t\to\infty$ limits , so, on strictly logical grounds, caution is needed. However, with the GP map $\lambda_t/t$ reaches  its limit value well before saturation is attained; moreover, it is the whole
$\lambda_t$ vs $t$ curves, and not only their slopes, that appear to stabilize at large $N$. This may be an empirical indication that, at least for the very smooth $\psi$, $\delta\psi$ which were used, at large $N$ the numerical
finite-N LEs do indeed mirror properties of the $\infty$-dimensional dynamics.

A measure  for the filling of the basis is provided by the normalized Inverse Participation
Ratio (IPR):
\begin{equation}
\label{ipr}
\IPR(\psi)\;=\;\left(N\sum\limits_{n=1}^N|\widehat{\psi}(n)|^4\right)^{-1}
\end{equation}
The maximum value of $\IPR(\psi)$ is $1$ and is attained
when $|{\hat\psi}(n)|=1/\sqrt{N}$, $\forall n$; and the average of $\IPR(\psi)$ over the uniform, normalized measure on the unit sphere in $\CM^N$ is asymptotically  equal to $0.5$ in the limit $N\to+\infty$\cite{zyc}.  Its dependence on $t$ is also shown in Fig.\ref{vstime}. Like energy, this quantity saturates, and remains thereafter quite close (within $0.06$\%) to the "microcanonical" average $0.5$. The microcanonical average of energy is $M(M+1)/6$, which, for $M=2^{17}$, is consistent with the value $0.5\log(E)\approx10.8$ that is observed at saturation in Fig.\ref{vstime}.

The approximate constancy in time (after saturation) of IPR  suggests ergodicity , because it would then be explained by  the concentration of measure (Levy's lemma) \cite{conc},  according to which IPR is close  to its  mean value\footnote{ Levy's lemma is applicable because the (non-normalized) IPR is a Lipschitz function on the hypersphere, with a Lipschitz constant independent of $N$.} in a subset of the sphere, whose measure is exponentially  close to $1$ for large $N$. Ergodicity (approximate, at least) is also supported by the observed independence of $\Lambda$ on the choice of an initial state (not so with $\Gamma$, which is sensitive to the filling of the basis by the chosen state).
If ergodicity of the $N$-dynamics is assumed, the following (non-exact) argument yields a good estimate for $\Lambda$ at large $N$. The state vector is now a $N$-vector
with components $\psi_{N,t}(n)=\psi_t(2\pi n/N)$, $(1\leq n\leq N)$, and lies on a hypersphere of radius $R_N$ in $\CM^N$, where
\begin{equation}
\label{norma}
\begin{split}
R_N^2\;&=\;\|\psi_{N,t}\|^2\;=\;\sum\limits_{n=1}^N|\psi_{N,t}(n)|^2\;\;\sim\\
&\sim\;\tfrac N{2\pi}\int_0^{2\pi}d\theta\;|\psi_t(\theta)|^2\;=\;
\tfrac N{2\pi}\;\;\mbox{\rm for}\;\;\; N\to\infty\;.
\end{split}
\end{equation}
The tangent dynamics is still described by eqn.(\ref{var}), where $\delta\psi_{N,t}$ is now a $N$-vector. A simple calculation yields:
\begin{equation}
\label{simple0}
\begin{split}
|\delta\psi_{N,t+1}(n)|^2\;&=\;(1\;+\;2\alpha^2|\psi_{N,t}(n)|^4)\;|\delta\psi_{N,t}(n)|^2\;+\\
&+\;2\mbox{\rm Im}\bigl((1+i\alpha\psi_{N,t}(n)^2\;\delta\psi_{N,t}(n)^2\bigr)\;.
\end{split}
\end{equation}
 For large $N$, ergodicity and Levy's lemma suggest replacing the $\psi_t$-dependent quantities in (\ref{simple0}) by their uniform averages $\langle .\rangle_{\mbox{\rm\tiny $R_N$}}$ over the hypersphere. This yields
\begin{equation}
\label{simple}
\begin{split}
|\delta\psi_{N,t+1}(n)|^2\;&=\;|\delta\psi_{N,t}(n)|^2\biggl(1\;+\;2\alpha^2\left\langle|\psi_{N,t}(n)|^4\right\rangle_{\mbox{\tiny $R_N$}}\;+\\
&+\;4\alpha\left\langle|\psi_{N,t}(n)|^2\right\rangle_{\mbox{\tiny ${R_N}$}}
\sin(2\phi_t(n))\biggr)\;,
\end{split}
\end{equation}
where $\phi_t(n)$ is the phase of $\delta\psi_{N,t}(n)$. In long-time iterates of (\ref{simple}), with rapid growth of $\delta\psi_{N,t}$ the dominant contribution can be assumed to come from the phase-independent term. Using that $\langle |\psi_{N,t}(n)|^4\rangle_{\mbox{\tiny $R_N$}}
=2R_N^4/N(N+1)\sim 2R_N^4/N^2$, and summing over $n$:
$$
\|\delta\psi_{N,t+1}\|^2\;\sim\;\|\delta\psi_{N,t}\|^2\;(1\;+\;4\alpha^2R^4_N/N^2)\;.
$$
Finally, for $N$ large (\ref{norma}) yields
\begin{equation}
\label{lyap}
\Lambda\;\approx\;\tfrac12 \log(1\;+\;\alpha^2/\pi^2)\;.
\end{equation}
Remarkably, $N$ has dropped out of this asymptotic formula.  The same  formula was obtained in ref.\cite{gc} for the rate of exponential growth of the square root of  energy , by means of a diffusion-based argument, and is in overall excellent agreement with numerical data shown in Fig.\ref{vsalpha}. Some agreement with numerical data for $\Lambda$ (but not with data  for $\Gamma$) persists for small values of $\alpha$,  when computations require
 much longer times than $t=300$. In the case when $\alpha=0.4$,  $\Lambda\approx 0.07$ was observed, whereas (\ref{lyap}) yields
$\Lambda=0.08$.\\
 The possible survival of stable regions at small $\alpha$ is beyond the scope of this paper. As shown in the next Section, for finite $N$  certain stationary orbits are linearly stable for $\alpha$ smaller than a threshold value, which however  decreases to $0$ in the limit $N\to\infty$, and
is already quite small for the cases considered here.
\begin{figure}
\begin{center}
\epsfxsize6cm\epsffile{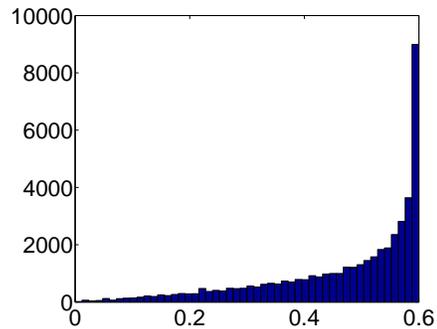}
\end{center}
\caption{\scriptsize Hystogram of positive Lyapunov exponents $\Lambda$ for the stationary state $u_0$ and $1\leq n\leq 2^{14}$; $\alpha=4$, $\tau=2\pi(\sqrt{5}-1)$.}
\label{hyst}
\end{figure}

\section{Stationary orbits.}
\label{sstat}
The Fourier basis functions $u_n(\theta)=(2\pi)^{-1/2}\exp(in\theta)$ $(n\in\ZM)$   satisfy
$$
\map(u_n)\;=\;e^{i(\bar{\alpha}-W_n)}\;u_n\;,
$$
where $\bar{\alpha}=\alpha/(2\pi)$, and $W_n=n^2\tau/2$.  They are therefore stationary states . If
$\bar\alpha=W_n+2k\pi$ for some $n,k\in\ZM$, they are also fixed points of the map on the unit sphere of $L^2(\TM)$ . Such values of $\bar\alpha$ are dense in $\RM$ whenever $\tau$ is incommensurate to $\pi$. For $\psi_0=u_r$  the variation equation (\ref{var}) takes a very simple form. Substituting $\delta\psi_t=\exp(i(\bar{\alpha}-W_r) t)\xi_t$,
\begin{equation}
\label{var2}
\xi_{t+1}\;=\;(1+i\bar\alpha)\;e^{iW_r} {\hat{\cal R}}_{\tau}\xi_t\;+\;2\pi i\bar\alpha\;e^{-iW_r}\;u^2_r\;({\hat{\cal R}}_{\tau}\xi_t)^*\;.
\end{equation}
Expanding $\xi_t$ over the Fourier basis : $\xi_t=\sum_{n\in\ZM}c_t(n)u_n$, and using that $2\pi u_n\;u^2_r=u_{2r+n}$,
\begin{equation}
\label{var3}
\begin{split}
c_{t+1}(n)\;&=\;e^{i(W_r-W_n)t}(1+i\bar\alpha)\;c_t(n)\;+\\
&+\;i\bar\alpha\;e^{-i(W_r-W_{2r-n})}c_t(2r-n)^*\;.
\end{split}
\end{equation}
Hence the subspace $\Sigma_{nr}$ spanned by $u_n,u_{2r-n}$ is invariant under the tangent dynamics . Denoting $d_t(n):=c_t(2r-n)^*$, for $n\neq r$
the dynamics in $\Sigma_{nr}$ are described by :
\begin{equation}
\label{mtx}
\begin{split}
\begin{vmatrix}
  c_{t+1}(n) \\
  d_{t+1}(n) \\
\end{vmatrix}
&=
\begin{vmatrix}
  (1+i\bar\alpha)\;e^{i(W_r-W_n)} & i\bar\alpha\;e^{-i(W_r-W_{2r-n})} \\
  -i\bar\alpha\;  e^{i(W_r-W_n)} & (1-i\bar\alpha)\;e^{-i(W_r-W_{2r-n})}\\
\end{vmatrix}
\times\\
&\times
\begin{vmatrix}
  c_t(n) \\
  d_t(n) \\
\end{vmatrix}
\;.
\end{split}
\end{equation}
while, for $\Sigma_{rr}$:
\begin{equation}
\label{l0}
c_{t+1}(r)\;=\;c_t(r)\;+\;2i\bar{\alpha}\;\text{\rm Re}\left(c_t(r)\right)
\;.
\end{equation}
 The matrix in eqn.(\ref{mtx}) is $t-$independent.
Denoting $L_{1,2}$ its eigenvalues one finds that $x_{1,2}:=
\exp(i(-W_{2r-n}+W_n)/2)L_{1,2}$ solve  the equation:
\begin{equation}
\label{char}
x^2\;-\;2\left[\cos\left(W_{r-n}\right)\;+\;{\bar\alpha}\sin\left(W_{r-n}\right)\right]\;x\;+\;1\;=\;0\;.
\end{equation}
For $m\in\ZM\setminus\{0\}$ denote $\alpha^{\pm}_m$ the least and the largest of the numbers $\tan(W_m/2)$ and $-\cot(W_m/2)$, if both numbers are finite; otherwise,  define  $\alpha^{\pm}_m=\pm\infty$. Then whenever $\alpha\notin B_{rn}:=[\alpha^-_m,\alpha^+_m]$,  eqn.(\ref{char}) has real roots, one of which is larger
than $1$ in absolute value , so it
yields a positive Lyapunov exponent $\Lambda_{rn}$, associated  to a direction in  the subspace $\Sigma_{nr}$:
\begin{gather}
\label{gvalue}
\Lambda_{rn}\;=\;\log\bigl(g_{r-n}+\sqrt{g_{r-n}^2-1}\bigr)\;,\nonumber\\
g_{l}\;=\;|\cos(W_l)\;+\;\bar\alpha\sin(W_l)|\;.
\end{gather}
Instead eqn.(\ref{l0}), rewritten as a map in $\RM^2$, has the single eigenvalue $1$ with algebraic multiplicity $2$ . It is thus marginally stable, with $|c_t(r)|\sim 2\bar\alpha t$ generically.

If $\tau$ is incommensurate to $\pi$, then the points $\exp(i(W_l))$, ($l\in\ZM$)  are dense and uniformly distributed in the unit circle, so  inf$\{\alpha^+_m, m\neq 0\}=0=$ sup$\{\alpha^-_m, m\neq 0\}$; hence, $\forall\alpha\neq 0$, and $\forall r\in\ZM$, condition
$\alpha\notin B_{rn}$ is satisfied for infinitely many values of $n$. For fixed $r$ and $\alpha\neq 0$, the relative frequency of such $n$ is $2\pi^{-1}\arctan(\bar\alpha)$, independently of the sign of $\alpha$ \footnote{This marks a sharp difference  with the  continuous-time GP equation in a ring \cite{berm}. The $u_n$ are  stationary states in that case, too, however none of them is unstable
as long as the GP coupling constant $g$ is positive. This also follows from the present analysis, because the GP equation is retrieved
from products of GP maps in the Lie-Trotter limit: $\alpha\to 0$, $\tau\to 0$, $\alpha=-g\tau$. In the $(\overline\alpha, W_{r-n})$ plane this limit corresponds to approaching the origin along the line $\overline\alpha=
-gW_{r-n}/(\pi(r-n)^2)$, which, for $g>0$ and small $\tau$ is entirely inside the stable region defined by condition $\overline\alpha\in  B_{rn}$. For $g<0$ it lies in the unstable region, provided that $|g|>\pi(r-n)^2/2$, consistently with results in ref.\cite{berm}.}.
Therefore, in the incommensurate case, and $\forall\alpha \neq 0$, each stationary state
$u_r$ is hyperbolic in infinitely many subspaces $\Sigma_{rn}$ , and elliptic in infinitely many of them. It is easy to compute
that
\begin{equation}
\label{suply}
\sup\limits_{n>0}\{\Lambda_{rn}\}\;=\;\log\left(|\bar\alpha|\;+\;\sqrt{1+\bar{\alpha}^2}\right)\;.
\end{equation}
Fig. \ref{hyst} shows a histogram of the positive LE for $r=0$ and $\alpha=4$. They cluster near the supremum (\ref{suply}), in this case equal to  $0.6$. The supremum (\ref{suply}) is always larger than the "ergodic" $\Lambda$ estimated by eqn.(\ref{lyap}), to which it is aymptotically equivalent in the limit $\alpha\to+\infty$.\\
In the commensurate case, when $\tau=4\pi P/Q$ with $P,Q$ mutually prime integers, the eigenvalues $e^{iE_n}$ take a finite number $Q$ of values . The same is then true of $\alpha^{\pm}_m$; moreover, if $Q$ is a prime number, then $\alpha^{\pm}_m$ do not vanish for $m\neq 0$, so they are bounded away from $0$ and infinity. Therefore for $\tau=4\pi P/Q$, $Q$ prime, $P$ prime to $Q$, there are finite values $\bar\alpha^{\pm}$ with
$\bar\alpha^{-}<0<\bar\alpha^{+}$ such that, in all subspaces $\Sigma_{nr}
$ with $n\neq r$,  all stationary orbits are linearly stable
if $\bar\alpha^{-}<\bar\alpha<\bar\alpha^+$, and are linearly unstable otherwise.
%%%%%%%%%%%%%%%%%%%%%%%%%%%%%%%%%%%%%%%%%%%%%%%%%%%%%%%%%%%%%%%%%%%%%%%%%%%%%%%%%%%%%%%%%%%%%%%%%%%%%%%%%%%%%%%%%%%%%%%%%%%%%%%%%%%%%%%%%%%%%%%%%%%
\begin{figure}
\begin{center}
\vskip 0.2cm
\epsfxsize8cm\epsffile{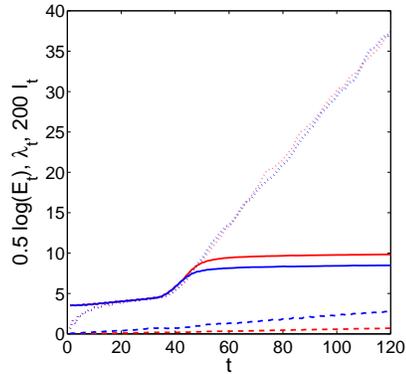}
\end{center}
\vskip 1cm
\caption{\scriptsize Same as Fig. \ref{vstime}, for $\tau=3\pi/2$, $\alpha=2.5$, and basis sizes $4^8$ (blue), $4^9$ (red).
}
\label{vstres}
\end{figure}
%%%%%%%%%%%%%%%%%%%%%%%%%%%%%%%%%%%%%%%%%%%%%%%%%%%%%%%%%%%%%%%%%%%%%%%%%%%%%%%%%%%%%%%%%%%%%%%%%%%%%%%%%%%%%%%%%%%%%%%%%%%%%%%%%%%%%%%%%%%%%%%%%%%

\section{"Resonant" case.}
\label{reso}
Although Golden Ratio incommensuration of $\tau$ to $2\pi$ was used in all hitherto shown numerical results, their
general features and the values of the maximal LEs appear to be essentially unchanged with less extreme incommensuration ({\it e.g}, with trascendental $\tau/(2\pi)=\pi$). However  the case when the period $\tau$ is commensurate to $2\pi$: $\tau=4\pi P/Q$, with $P,Q$ mutually prime integers, has significant differences. It  corresponds to the resonances of the quantum kicked rotor \cite{kr,ig}, and the special cases $\tau=4\pi$ and $\tau=2\pi$ are explicitly solvable for the QKR and for the GP map as well. In the latter cases the solutions are:
\begin{equation}
\begin{split}
\psi_t^{(4\pi)}(\theta) \;&=\;e^{i\alpha t|\psi_0(\theta)|^2}\psi_0(\theta)\;,\\
&\psi_{2t}^{(2\pi)}(\theta)\;=\;e^{i\alpha(t|\psi_0(\theta+\pi)|^2+
t|\psi(\theta)|^2)}\psi_0(\theta)\;.
\end{split}
\end{equation}
The tangent dynamics for the fundamental resonance $\tau=4\pi$ is solved by
\begin{equation}
\begin{split}
\delta\psi_t(\theta)\;=\;&e^{
i\alpha t|\psi_0(\theta)|^2}\;\bigl(\delta\psi_0(\theta)\;+\\
&2i\alpha\; t\;\psi_0\;{\mbox{\rm Re}}(\psi_0(\theta)^*
\delta\psi_0(\theta))\;
\end{split}
\end{equation}
so $\|\delta\psi_t\|$ increases at most linearly with $t$, and all LE vanish. The solution for $\tau=2\pi$ has a slightly more complicated form,  but the conclusion is the same.
\begin{figure}
\epsfxsize8cm\epsffile{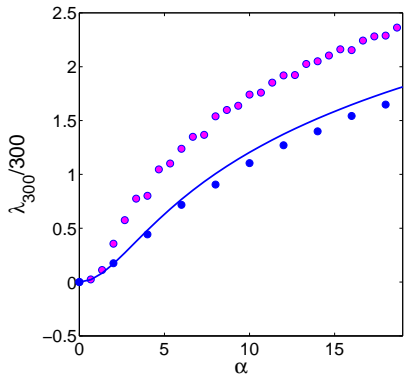}
%\vskip1cm
\caption{\scriptsize Finite time Lyapunov exponent $\lambda_t/t$ vs $\alpha$ for $t=300$. Full line, eqn.(\ref{lyap}); magenta circles, $\tau=8\pi/3$, basis size $3^9$; blue circles, $\tau=14\pi/11$, basis size $11^5$.}
\label{vsalphares}
\end{figure}
If  $Q$ is an arbitrary integer, every $\psi\in L^2([0,2\pi/Q])$ can be rewritten as the vector-valued function  $\vec\psi_{\theta}$ that, for $\theta\in[0,2\pi/Q]$, has components $\vec{\psi}_{\theta}(j)=\psi(\theta+2\pi (j-1)/Q)$, $(j=1,\ldots,Q)$; this vector will be termed the fiber of $\psi$ at the point $\theta$. Then:
\begin{equation}
\label{fibnorm}
\|\psi\|^2\;=\;\int_{(0,2\pi/Q)}d\theta\;\|{\vec\psi}_{\theta}\|^2\;,
\end{equation}
where the norm under the integral sign is the $\CM^Q$ one. Therefore $L^2(\TM)$ can be identified with the Hilbert space  $L^2([0,2\pi/Q])\otimes\CM^Q$. If $\tau=4\pi P/Q$, then $\map$ "acts fiberwise", that is:
\begin{equation}
\label{fib1}
\map\;=\;\mathbb I\otimes\mapth,
\end{equation}
where $\mathbb I$ is identity, and $\mapth$ is a nonlinear map in $\CM^Q$ such that, for $\Psi\in\CM^Q$,
\begin{gather}
\label{eqres}
\mapth(\Psi)(j)\;=\;e^{i\alpha\hat G\Psi(j)^*\hat G\Psi(j)}\;\hat G\Psi(j)\;, \;\;\;1\leq j,k\leq Q\;.
\end{gather}
where the linear unitary operator $\hat G$ is described by the matrix:
\begin{gather}
G(j,k)\;=\;\tfrac1Q\sum\limits_{s=0}^{Q-1}\;e^{2\pi is(j-k)/Q}\;a_{s+1}\;,\nonumber\\
a_r\;=\;e^{-iW_r}\;=\;e^{-2i\pi P(r-1)^2/Q}\;,\;\;\;\;1\leq r \leq Q\;.
\end{gather}
Therefore, in the commensurate case the GP map has infinitely many conserved quantities, namely $\|{\vec\psi}_{\theta}\|$, $0\leq\theta\leq 2\pi/Q$ or, equivalently, the countably many quantities $I_n(\psi)=( \psi, \exp(inQ\theta)\psi)$, ($n\in\ZM$).
The fiber map (\ref{eqres}) does not depend on $\theta_0$. However it preserves the $\CM^Q$ norm and so it is conveniently studied by restricting to the unit sphere in $\CM^Q$. In the following, $\overline{U}_{\alpha}$ denotes the restriction of (\ref{eqres}) to the unit $\CM^Q$ sphere, so eqn.(\ref{fib1}) rewrites in the form:
\begin{equation}
\label{fib2}
 \overrightarrow{\map(\psi)}_{\theta}\;=\;\|{\vec\psi}_{\theta}\|\;\bar U_{\alpha(\theta)}\left({\vec\psi}_{\theta}/\|{\vec\psi}_{\theta}\|\right)\;,
\end{equation}
where:
\begin{equation}
\label{alfateta}
\alpha(\theta)\;=\;\alpha\;\|{\vec\psi}_{\theta}\|^2\;.
\end{equation}
The map $\bar U_{\alpha}$ will be termed "fiber map" in the following. It is symplectic and defines a classical dynamical system in the unit sphere in $\CM^Q$ . Existence of LEs is then granted. The variation equation around a reference orbit $\{\bar U_{\alpha}^t(\Psi_0)\}$ is obtained from eqn.(\ref{var}) on replacing
$\psi(\theta),\delta\psi(\theta)$ by ${\vec\psi}(j),\overrightarrow{\delta\psi}(j)$, and $\hat{\cal R}$ by the matrix $G$.
\begin{figure}
\epsfxsize8cm\epsffile{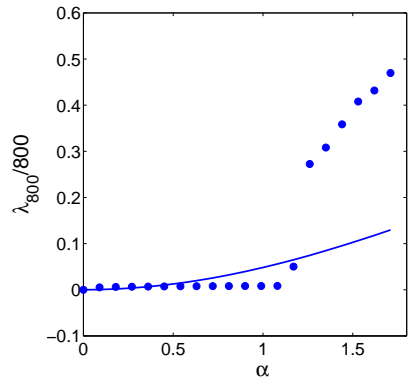}
\caption{\scriptsize The small-$\alpha$ region of Fig.\ref{vsalphares}, $\tau=8\pi/3$, $t=800$.}
\label{vsalphares1}
\end{figure}
Turning to numerics:
Fig.\ref{vstres} was constructed like Fig.\ref{vstime}, except that now $\tau=3\pi/2$, and it looks qualitatively similar. The main difference is that the onset of exponential growth, both for $\lambda_t$ and for energy, occurs significantly later, and is preceded by a relatively long linear growth; and the increase of IPR is much slower.  Fig.\ref{vsalphares} shows the dependence of the finite-time LE $\lambda_t/t$ at $t=300$ vs $\alpha$. While in the incommensurate case (Fig.\ref{vsalpha}) this dependence follows the theoretical law (\ref{lyap}), it strongly deviates for low-$Q$ resonances.
Such differences are due to the presence of the above mentioned constants of motion $I_n(\psi)$. These in particular imply, that the finite-dimensional reduced dynamics is not ergodic; this invalidates the underlying argument of
 formula (\ref{lyap}, and
suggests that the observed LEs should not be independent of the choice of a trajectory. From eqs.(\ref{fibnorm}), (\ref{fib2}),(\ref{alfateta}) it is seen that stability of the global $L^2(\TM)$ dynamics is fully  governed by the stability of the fiber maps; in particular, for smooth initial $\psi_0$,  the maximal LE is expected to coincide with that of the fiber map $\overline{U_{\alpha_0}}$, where $\alpha_0=$max$\{\alpha(\theta), 0\leq\theta\leq 2\pi/Q\}$. \\
Numerical data suggest an integrable-to-chaotic transition of the fiber map on increasing $\alpha$. The statistical dispersion over the unit sphere of the time-average $T^{-1}\sum_1^T \IPR(\overline{U_{\alpha}}^t\Psi_0)$ at large fixed $T$ (as a function of $\Psi_0$)  is seen to sharply decrease on increasing  $\alpha$, indicating that ensemble-averages  and time-averages of the function $\IPR(\Psi)$ tend to coincide. Along with the small dispersion of LEs of randomly chosen trajectories (for $\alpha=4$, $t=300$, $\tau=8\pi/3$  the standard deviation of $\lambda_t$ over an ensemble of $500$ randomly generated orbits
 is less that $2\%$), this provides empirical support for at least approximate ergodicity. If so, then stronger ergodic properties may be conjectured, on account of the positivity of LEs. Note that ergodicity of fiber dynamics is in no  contradiction to  the constants  $I_n(\psi)$, because these  are constant on fibers.
 LEs of the finite-dimensional fiber map can be expected to follow formula (\ref{lyap}), provided $Q$ is sufficiently large, and $\alpha$, too,  is sufficiently large.
Indeed, the rhs of fig.\ref{vsalphares} suggests that LE may vanish at small $\alpha$. The fiber dynamics has $Q$ stationary states, for which the analysis in Sect.\ref{sstat} still applies, provided all integer indices $n,r$ are taken mod$(Q)$. At small $Q$, the threshold value $\alpha_{\mbox{\it\tiny cr}}$,
below which such stationary trajectories are linearly stable, is not any more negligible (see remarks in the end of sect.\ref{sstat}) as it was in the incommensurate case, and this fact may be responsible for stable islands for small $\alpha$, both in the fiber and in the global dynamics.

\section{Concluding remarks.}
\label{concr}
In this paper the GP map was studied as a classical dynamical system. However the GP equation, of which it is a byproduct,  is a quantum construct, that was devised to model the effect of interactions in dilute Bose-Einstein condensates. The question may then be asked, if  chaoticity of the GP map may be taken  as an instance of genuine chaotic behaviour in quantum mechanics.
However, nonlinear deterministic Schr\"odinger equations are not expected to preserve, on a fundamental level, the basic distinctive features of quantum mechanics \cite{gisrig}; and dynamical chaos is indeed a fundamental issue.
 Nevertheless the GP equation is an efficient  mean-field approximation for a many body quantum dynamics, and it would be very interesting to know whether and in which form  the exponential instability of the GP map may be mirrored in an exact many-body dynamics.
 \vskip0.5cm\noindent
{\bf Acknowledgment} The present Author is indebted to the Authors of ref.\cite{gc} for communicating their results before publication.

\end{document}